\documentclass[aps,prb,twocolumn,amsmath,amssymb,nolongbibliography,superscriptaddress]{revtex4-2}
\usepackage{graphicx}
\usepackage{xcolor}
\usepackage{upgreek}
\usepackage{hyperref}
\usepackage[normalem]{ulem}

\hypersetup{
	colorlinks=true, 
	citecolor=blue, 
	linkcolor=blue, 
	urlcolor=blue, 
}
\DeclareMathOperator{\artanh}{artanh}
\DeclareMathOperator{\sech}{sech}

\begin{document} 
\newcommand{\reffig}[1]{Fig.\,\ref{#1}}
\newcommand{\refeq}[1]{Eq.\,(\ref{#1})}
\newcommand\numberthis{\stepcounter{equation}{1}\tag{\theequation}}
\newcommand{\change}[1]{{\color{blue} #1}}
\newcommand{\remove}[1]{\textcolor{red}{\sout{#1}}}

\title{Thermal conductance and non-equilibrium superconductivity in a diffusive NSN wire probed by shot noise}

\author{A.V.~Bubis} 
\affiliation{Skolkovo Institute of Science and Technology, Nobel street 3, 121205 Moscow, Russian Federation}
\affiliation{Institute of Solid State Physics, Russian Academy of Sciences, 142432 Chernogolovka, Russian Federation}
\author{E.V.~Shpagina} 
\affiliation{Institute of Solid State Physics, Russian Academy of Sciences, 142432 Chernogolovka, Russian Federation}
\affiliation{National Research University Higher School of Economics, 20 Myasnitskaya Street, Moscow 101000, Russia}
\author{A.G.~Nasibulin}
\affiliation{Skolkovo Institute of Science and Technology, Nobel street 3, 121205 Moscow, Russian Federation}
\affiliation{Aalto University, P. O. Box 16100, 00076 Aalto, Finland}
\author{V.S.~Khrapai} 
\affiliation{Institute of Solid State Physics, Russian Academy of Sciences, 142432 Chernogolovka, Russian Federation}
\affiliation{National Research University Higher School of Economics, 20 Myasnitskaya Street, Moscow 101000, Russia}
\begin{abstract}
We investigate diffusive nanowire-based structures with two normal terminals on the sides and a central superconducting island in the middle, which is either grounded or floating. Using a semiclassical calculation we demonstrate that both device layouts permit a quantitative measurement of the energy-dependent sub-gap thermal conductance $G_\mathrm{th}$ from the spectral density of the current noise. In the floating case this goal is achieved without the need to contact the superconductor provided the device is asymmetric, that may be attractive from the experimental point of view. In addition, we observe that the shot noise in the floating case is sensitive to a well-known effect of non-equilibrium suppression and bistability of the superconducting gap. Our calculations are directly applicable to the multi-mode case and can serve as a starting point to understand the shot noise response in open one dimensional Majorana device.
\end{abstract}

\maketitle

\section{Introduction}

Electronic transport in hybrid semiconductor-superconductor devices is getting a second breath in the context of recent topological band theory. One of the promising directions is a realization of topological superconductivity in a proximitized semiconducting nanowire (NW)~\cite{Lutchyn2010, Oreg2010}, accompanied by emerging Majorana zero modes (MZMs) localized at its ends~\cite{Alicea2012}. While all the prerequisites for this noble goal are there, including ballistic single-mode transport~\cite{Gul2018}, strong spin-orbit coupling~\cite{Mourik2012} and thin superconducting shell capable to withstand strong magnetic fields~\cite{Krogstrup2015}, the non-local character of the proposed MZMs remains to be proved. 

The MZMs non-locality can be probed with nonlocal conductance measurements in normal-superconductor-normal (NSN) NW devices~\cite{Menard2020, Puglia2021, Yu2021}. Such a three-terminal setup approach allows to overcome the problem of superconducting shell shunting the quasiparticle charge transport and can capture the MZMs via end-to-end conductance correlations~\cite{Lai2019} and Andreev rectification effect~\cite{Rosdahl2018}. Alternative to charge transport are nonlocal thermal conductance and shot noise measurements, which provide a universal signature of the topological phase transition even in presence of a moderate disorder~\cite{Akhmerov2011}. At further increasing the disorder, the thermal conductance becomes the only measure of the non-local quasiparticle response~\cite{Pan2021}. In the absence of heat transfer through the superconducting shell, one can expect the thermal conductance to be informative also in NSN devices with a floating S-island~\cite{Fu2010}. Possible relation to the shot noise measurements in such structures remains, however, unknown~\cite{Ulrich2015}.

A correspondence between the shot noise and thermal conductance is a generic effect not limited to the Majorana case. A doubling of the shot noise in disordered NS junctions~\cite{Jehl2000, Kozhevnikov2000} is fundamentally related to the suppressed heat transfer in the S-lead~\cite{Nagaev2001, Nagaev2001b}, and can be  useful to probe the sub-gap density of states in such structures~\cite{Tikhonov2016, Sahu2019}. In NSN NW-based devices the shot noise and thermal conductance are directly related in the limit of charge neutral quasiparticle transport, that was demonstrated in a recent experiment set up in a trivial superconducting phase~\cite{Denisov2021}. It is instructive to trace the interplay of disorder scattering and Andreev reflection in the framework of semiclassical multi-mode NSN devices. By mixing quasiparticle trajectories traversing the proximity region at different angles the disorder randomizes the number of Andreev reflections (ARs) of a sub-gap quasiparticle from the superconducting shell~\cite{Kopnin2004, Denisov2021}. Since each AR process inverts the quasiparticle charge~\cite{Andreev1964}, statistically this favors the charge neutrality of the quasiparticle population. In addition, moderate disorder may enhance the heat conductance by promoting the escape of retro-reflected quasiparticles from poorly propagating trajectories~\cite{Andreev1965, Kopnin2004}. Thus diffusive multi-mode  NSN structures represent a perfect test bed of the relation between the shot noise and thermal conductance.

Semiconductor-based hybrids also offer unique possibilities for the investigation of non-equilibrium effects caused by quasiparticles in the superconductor. Widely explored in all-metal NSN structures~\cite{Keizer2006,Snyman_2009,Catelani_2010,Vercruyssen_2012,Serbyn_2013}, in modern NSN nanowire devices this direction has not yet received the attention it deserves, being strongly outweighed by the MZM research. With semiconductors at hand, the device asymmetry, determined by the conductances of the N segments, can be tuned by gate voltages that allows additional control over the shape of a non-equilibrium quasiparticle energy distribution in the floating S-island, and thus over the value of the superconducting gap~\cite{Snyman_2009}. This is in contrast to a fixed asymmetry in all-metal devices equipped with tunnel barriers~\cite{Snyman_2009,Catelani_2010}. To our best knowledge, the manifestation of this kind of non-equilibrium effects in shot noise has not been investigated so far.

Here, we investigate a diffusive NSN NW-based device from the perspective of shot noise measurements using  a semiclassical approach of Nagaev and B\"{u}ttiker~\cite{Nagaev2001, Nagaev2001b}. We also discuss how the energy dependent sub-gap thermal conductance of the central S segment owing to proximity effect can be deduced from the Usadel theory~\cite{Keizer2006} and how the non-equilibrium effects impact the superconducting gap~\cite{Snyman_2009}. We consider two different layouts widely used in Majorana setups with a central superconducting island either connected to a grounded macroscopic terminal or floating. Grounding the superconductor turns it into a perfect sink for charge and for above-gap quasiparticles, that is a crucial distinction between these cases. In the grounded case the nonlocal shot noise is sensitive to the thermal conductance of the proximitized segment. In the floating case the impact of the thermal conductance on the shot noise is weaker and depends on the device asymmetry. This makes the floating island geometry considerable for experimentalists, since a technically challenging step of contacting the thin superconducting shell~\cite{Krogstrup2015, Gazibegovic2017, Krizek2018b, Vaitiekenas2018} can be omitted in this case. On the other hand, a peculiar behaviour of the superconducting gap on the bias voltage and asymmetry in the floating geometry opens up a new avenue for the investigation of non-equilibrium effects in proximity devices by means of the shot noise measurements.

\section{Semiclassical model}

\begin{figure}[b]
	\begin{center}
		\includegraphics[width=1\columnwidth]{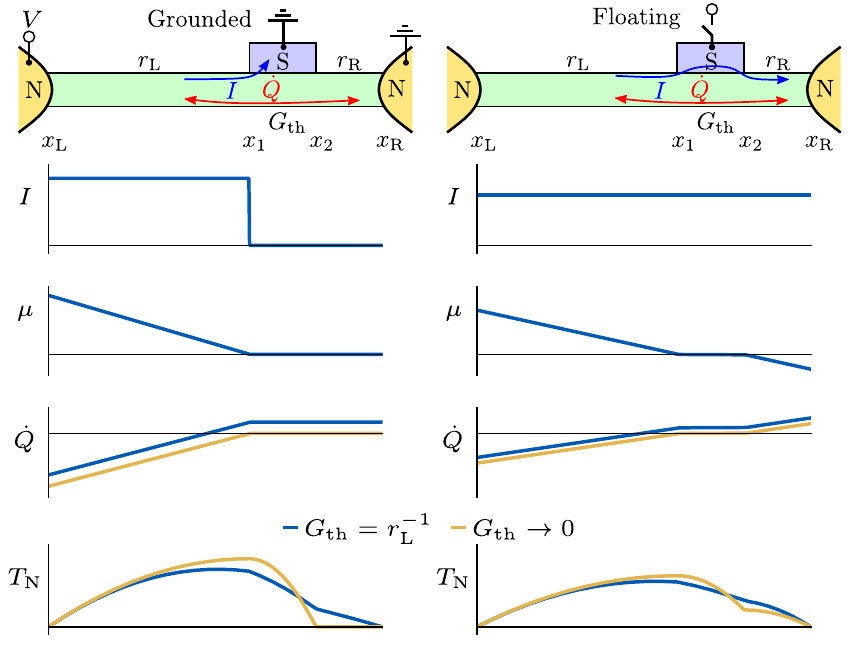}
	\end{center}
	\caption{Schematics of the two NSN device configurations and coordinate dependencies of the key physical quantities: $I$ -- electrical current, $\mu$ -- local chemical potential,   $\dot{Q}$ -- heat flux integrated over energy, $T_\mathrm{N}$ -- local noise temperature. The boundaries of the S segment are marked by vertical dashed lines. The parameters used for calculations are $r_\mathrm{L} = 6\,\mathrm{k\Omega}$, $r_\mathrm{R} = 2\,\mathrm{k\Omega}$, $T = 0$, $|eV| < \Delta$.}
	\label{fig1}
\end{figure}

We consider a diffusive nanowire (NW) connected to two normal reservoirs (N) and a superconducting contact in the middle (S), which divides the wire in two normal sections with the resistances $r_\mathrm{L}$ and $r_\mathrm{R}$ (see the first row in \reffig{fig1}). The normal NW segments on both sides of the S contact are assumed to be much longer than the superconducting coherence length and the applied voltage is much higher than the Thouless energy, $L \gg \sqrt{\hbar D/eV},\sqrt{\hbar D/\Delta}$, where $D$ is the diffusion coefficient. This allows to neglect the penetration of the superconducting condensate from the proximity region underneath the S contact into the normal segments and treat them as metallic diffusive conductors~\cite{Nagaev2001, Nagaev2001b}. The length of the S segment (the part of the device consisting of a part of NW and S contact above it) is assumed to be much larger than both the NW diameter and the superconducting coherence length, which enables us to describe the quasiparticle transport via this segment as effectively one-dimensional and neglect the processes of Cooper pair splitting and elastic cotunneling~\cite{Kirsanov2019}, as well as the Coulomb blockade effects~\cite{Fu2010}. S/NW interface quality is assumed to be nearly ideal, so that the probability of the AR by far exceeds that of the normal quasiparticle reflection. Inelastic scattering in the NW and in the S-island is absent. The thermal conductance $G_\mathrm{th}$ of the S segment is assumed to be finite at energies below the superconducting gap $\Delta$. Above the gap, the S contact shunts both electrical and thermal currents, thus in this energy range the thermal conductance is much larger than that of the adjacent normal NW segments and assumed to be infinite. Throughout the paper we define the thermal conductance similarly to the electrical conductance as $G_\mathrm{th} \equiv e^2\nu^*\,D^* A/L$, where $e$ is the elementary charge and $\nu^*,\,D^*$ are, respectively, the density of states and diffusion coefficient of the sub-gap quasiparticles in the S segment, $A$ is the cross-section of the NW and $L$ is the length of the S segment. This choice is convenient for our purpose of solving a non-equilibrium finite-bias problem. Note that in general $G_\mathrm{th}$ may depend on energy owing to the superconducting proximity effect, as we address  later below.

In the following we consider two possible realizations of the NSN devices (see schematic device configurations in Fig.~\ref{fig1}), with the central S segment either being a part of a grounded superconducting terminal (reservoir) or a floating island. These layouts are referred to, respectively, as the Grounded and the Floating cases. The Grounded case corresponds to a three-terminal NSN device for which the bias voltage $V$ is applied to the left N terminal, whereas the S terminal and the right N terminal are grounded. Corresponding chemical potentials are $\mu_\mathrm{L} = -eV$, $\mu_\mathrm{S} = 0$ and $\mu_\mathrm{R} = 0$. The Floating case corresponds to a two-terminal NSN device for which, without the loss of generality, we also choose the chemical potential of the S segment equal to zero $\mu_\mathrm{S} = 0$, therefore $\mu_\mathrm{L} = -eVr_\mathrm{L}/(r_\mathrm{L}+r_\mathrm{R})$ and  $\mu_\mathrm{R} = eVr_\mathrm{R}/(r_\mathrm{L}+r_\mathrm{R})$.

\section{General considerations}

Before going into the details of our calculation, we illustrate the underlying physics of the NSN structure in a few representative cases at a zero bath temperature and finite bias in Fig.~\ref{fig1}, neglecting so far the proximity effect and non-equilibrium superconductivity effects in the S segment~\cite{Keizer2006}.  Blue arrows in the upper sketches indicate the propagation of the electric current~$I$. The key difference between both device configurations is that the electric current~$I$ flows only through the left NS segment in the Grounded case and through the whole device in the Floating case. Note that in the latter case the current in the S segment is carried by the Cooper condensate and flows predominantly inside the superconductor. The propagation of current has obvious consequences for the local chemical potential $\mu$, see the spatial profiles of both quantities in respective panels of Fig.~\ref{fig1}. This and other data are the results of the calculations with the parameters mentioned in the caption. Non-equilibrium sub-gap quasiparticles gain energy from the electric field and propagate diffusively along the NW, relaxing in one of the N terminals. The direction of the heat flux is indicated by the red arrows. Finite $G_\mathrm{th}$ enables non-equilibrium quasiparticles to traverse the S segment and results in a non-zero heat flux $\dot{Q}$ in the S segment. The spatial dependence of the heat flux is shown in the corresponding panel of Fig.~\ref{fig1}. Note that in general $\dot{Q}$ changes sign somewhere in the middle of the NSN device and depends on coordinate, which is a consequence of the Joule heating released in the normal segments that relaxes in the N terminals. In the limit of $G_\mathrm{th}=0$ shown by yellow lines, as well as in the special case of symmetric Floating NSN device, the heat flux via the S segment vanishes and the two normal segments completely decouple. Non-equilibrium quasiparticle populations which build up in the biased NSN NW are characterized by coordinate-dependent electronic energy distributions (EEDs) $f(\varepsilon)$, which we calculate in the next section. Relevant to the shot noise measurements in diffusive conductors is the notion of the noise temperature  $T_\mathrm{N}= (k_\mathrm{B})^{-1} \int f(1-f) d \varepsilon$, which is a measure of local non-equilibrium. The lower panels of Fig.~\ref{fig1} demonstrate the spatial profiles of the $T_\mathrm{N}$. Note that in the limit of $G_\mathrm{th}=0$ the right N segment remains in equilibrium in the Grounded case, whereas it acquires a finite $T_\mathrm{N}>0$ in the Floating case.

\section{Energy distributions\label{section_EED}}

Following the semiclassical approach of Nagaev and B\"{u}ttiker~\cite{Nagaev2001, Nagaev2001b} we calculate the EEDs $f_1(\varepsilon)$ and $f_2(\varepsilon)$ on the two boundaries of the S segment at $x=x_1$ and $x=x_2$, see Fig.~\ref{fig1}. The vicinity of the superconductor imposes an important constraint on the EED at sub-gap energies. For the case of a perfect lateral interface between the superconductor and the NW, which is assumed below, AR is the only process of quasiparticle scattering from the interface. Since the number of AR in the diffusive case is a random quantity~\cite{Denisov2021}, statistically this results in equal amounts of the electron-like and hole-like sub-gap quasiparticles in the S segment, that is the function $f(\varepsilon, x)$ for $x \in [x_1, x_2]$ obeys the relation $f(\varepsilon, x) = 1 - f(-\varepsilon, x)$. Note that this symmetry automatically guarantees that the chemical potential of the sub-gap quasiparticles coincides with that of the Cooper pairs, i.e. $\mu(x)=0$.

At sub-gap energies the electric current in the S segment is carried by the Cooper condensate, therefore the conservation of electric current cannot be used to set the boundary conditions for the EED. The proper boundary conditions for $|\varepsilon|\leq\Delta$ are obtained from the conservation of a partial heat flux at a given energy. We define such a heat flux as $\delta\dot{Q}(\varepsilon, x)=-\nu(\varepsilon)D(\varepsilon)[\varepsilon-\mu(x)]\nabla f(\varepsilon, x)\delta\varepsilon$, where $\nu(\varepsilon)$ and $D(\varepsilon)$ are the density of states, diffusion coefficient and chemical potential at this energy and $\delta\varepsilon$ is the width of infinitesimal energy window. Having in mind that AR mixes the two types of quasiparticles in our hybrid system, we observe that the correct conserved quantity the sum of the partial heat fluxes carried by the electron-like and hole-like quasiparticles at the same excitation energy $|\varepsilon|$. That is the quantity $\delta\dot{Q}(\varepsilon, x) + \delta\dot{Q}(-\varepsilon, x)\propto-\nabla F(\varepsilon, x)$, where we introduced $F(\varepsilon,x) \equiv f(\varepsilon,x) - f(-\varepsilon,x)$.

In the diffusive transport regime, within each NW segment the EED satisfies the equation $\frac{\partial^2}{\partial x^2} f(\varepsilon, x)=0$ and interpolates linearly as a function of $x$ between the boundary conditions~\cite{Nagaev1992}. Thus, the conservation of the heat flux is expressed as:
\begin{gather*} 
	|\varepsilon| < \Delta:\\
	\begin{align}
		\begin{split}
		\frac{F_1(\varepsilon) - F_\mathrm{L}(\varepsilon)}{r_\mathrm{L}} &= 
		(F_2(\varepsilon) - F_1(\varepsilon)) \cdot G_\mathrm{th} \\
		\frac{F_\mathrm{R}(\varepsilon) - F_2(\varepsilon)}{r_\mathrm{R}} &= 
		(F_2(\varepsilon) - F_1(\varepsilon)) \cdot G_\mathrm{th},
		\label{eq:Fs}
		\end{split}
	\end{align}	
\end{gather*}
where the functions $F_\mathrm{L,\,R}$ are given by the equilibrium Fermi-Dirac EEDs $f_0(\varepsilon-\mu_\mathrm{L,\,R})$ and functions $F_\mathrm{1,\,2}$ correspond to the boundaries of the S segment at $x=x_1$ and $x=x_2$. The solution of \refeq{eq:Fs} is straightforward:
\begin{gather*} 
	|\varepsilon| < \Delta:\\
	\begin{align}
		\begin{split}
		f_1 &= \frac{1}{2}\left(1 + \frac{F_\mathrm{R}r_\mathrm{L} + F_\mathrm{L}(r_\mathrm{R} + 1/G_\mathrm{th})}{r_\mathrm{L} + 1/G_\mathrm{th} + r_\mathrm{R}}\right) \\
		f_2 &= \frac{1}{2}\left(1 + \frac{F_\mathrm{L}r_\mathrm{R} + F_\mathrm{R}(r_\mathrm{L} + 1/G_\mathrm{th})}{r_\mathrm{L} + 1/G_\mathrm{th} + r_\mathrm{R}}\right).
		\label{eq:f_subgap}
		\end{split}
	\end{align}
\end{gather*}

For above-gap energies $|\varepsilon|>\Delta$ we neglect the ARs and assume that the S segment essentially behaves as a piece of normal metal with the conductance much higher than that of the normal NW segments. Thus $f(\varepsilon)$ is independent of coordinate for $x \in [x_1, x_2]$. In the Grounded case  the EED in the S segment is simply given by the equilibrium EED with $\mu = \mu_\mathrm{S} \equiv0$. In the Floating case the EED is calculated from matching the quasiparticle currents in the neighboring normal segments at $x=x_1$ and $x=x_2$ and acquires a familiar linear combination of the boundary conditions at the left and right N terminals~\cite{Nagaev1992}:
\begin{gather*}
	|\varepsilon| \ge \Delta:\\
	\begin{align}
		f_{1,2} &= 
		\begin{cases}
		f_0, & \text{Grounded}\\
		(f_\mathrm{R}r_\mathrm{L} + f_\mathrm{L}r_\mathrm{R})/(r_\mathrm{L}+r_\mathrm{R}), & \text{Floating}
		\end{cases}
	\label{eq:f_abovegap}
	\end{align}
\end{gather*}

\begin{figure}[t]
	\begin{center}
		\includegraphics[width=1\columnwidth]{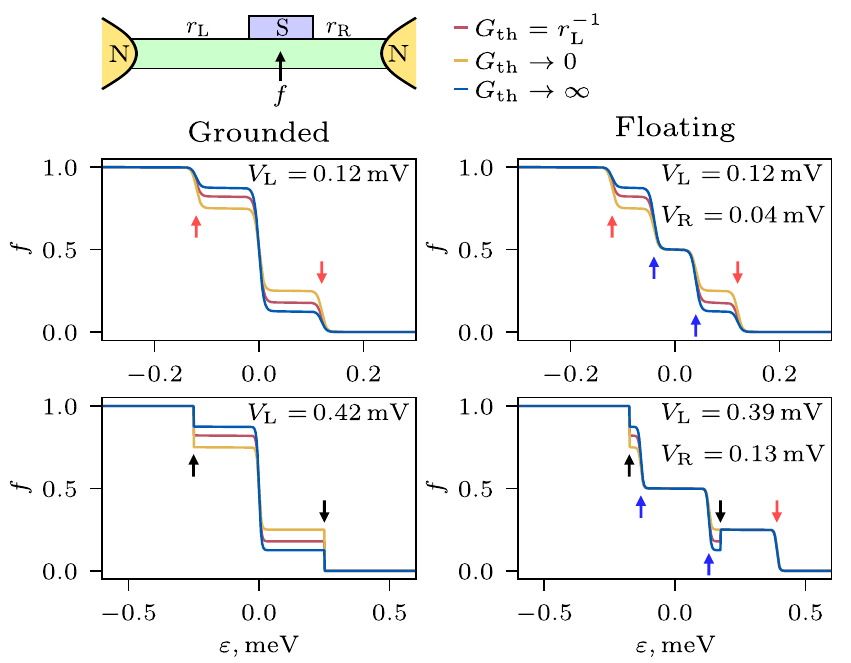}
	\end{center}
	\caption{The EED in middle of the S segment, $f\equiv(f_1+f_2)/2$,  at several applied biases. Arrows on the graphs denote specific energies: black arrow -- $\Delta$, red arrow -- $|eV_\mathrm{L}|$, blue arrow -- $|eV_\mathrm{R}|$. Resistances of the normal segments are the same as in \reffig{fig1}, $T = 50\,\mathrm{mK}$.}
	\label{fig2}
\end{figure}

It is straightforward to see that the Eqs.~(\ref{eq:f_subgap})~and~(\ref{eq:f_abovegap}) contain a limiting case of a diffusive wire with N and S contacts. This situation is achieved for both device layouts taking $r_\mathrm{R} = 0$ and  $G_\mathrm{th} = 0$. In this case the voltage drop on the right  N-section is zero, $\mu_\mathrm{R} = 0$, and $f_2(\varepsilon)$ is given by the equilibrium EED. The energy distribution at $x=x_1$ has a well-known double-step shape~\cite{Nagaev2001} with $f_1(\varepsilon) = 1/2$  for $|\varepsilon| < \Delta,\, |eV|$ and 0 or 1 otherwise.

In general, the situation is more sophisticated and in order to demonstrate the main physics we discuss now the EED for a marginal case of the energy independent $G_\mathrm{th}$. This situation corresponds to the case of an Andreev wire with the diameter much larger than $\xi_n=\sqrt{\hbar D/k_BT}$ that allows to neglect the proximity effect at all relevant energies.  In \reffig{fig2} we plot the EED in the middle of S segment for both device layouts at various bias voltages. Different line styles correspond to three representative values of $G_\mathrm{th}$, see the legend. The vertical arrows show the positions of the chemical potentials and the gap edges as explained in the caption. For small enough bias voltages the EED has three steps in the Grounded case and four steps in the Floating case (upper panels in both columns). These steps are smeared by the finite temperature. The effect of the increasing $G_\mathrm{th}$ is to diminish the amount of sub-gap quasiparticles in the S segment by sinking them into the right N terminal. This results in the increase/decrease of the EED for $\varepsilon<0$/$\varepsilon>0$. In addition, the step in $f(\varepsilon)$ occurs each time the voltage drop on one of the normal NW segments equals $\Delta$, these steps at the gap edges are sharp. In the Grounded case $f(|\varepsilon|>\Delta)$ is equilibrium, see the Eq.~(\ref{eq:f_abovegap}) and the lower panel in the left column of \reffig{fig2}. In the Floating case the situation is much richer and the EED may be non-monotonic and exhibit up to five steps depending on the relation between the voltage drops $V_\mathrm{L,\,R}$ and the superconducting gap, see the lower panel in the right column of \reffig{fig2}. In this panel, we have taken into account a renormalization of $\Delta$ caused by the non-equilibrium quasiparticle EED in the superconductor calculated for the sweep-up solution branch, see section~\ref{Usadel} for the details. Note that such EEDs in the S segment can be directly measured with a local tunnel probe using transport~\cite{Pothier1997} or noise~\cite{Tikhonov2020} approaches.

\section{Relation to a complete theory}

Full understanding of a non-equilibrium configuration involving superconductivity usually requires a self-consistent numerical solution of the Usadel equations~\cite{kopnin2001theory}. Such an approach was realized in Ref.~\cite{Keizer2006} for the all-metal NSN structure, that has certain similarities with our geometry. The key difference is that in NW-based devices considered here the voltage drop on the superconductor is negligible and the shape of the non-equilibrium EED is controlled by the resistances of the N segments and thermal conductance of the proximity region, as discussed in the previous section~\ref{section_EED}. Below  we briefly outline how our semiclassical model is related to a complete solution of  the Usadel equations. For the sub-gap quasiparticles we qualitatively discuss the proximity effect induced modification of the heat transport in the normal core of the S segment, that is relevant for both the Grounded and the Floating device layouts. In addition, in the Floating case the superconductor is decoupled from the reservoir and the non-equilibrium EED has a crucial impact on the superconducting gap. Below we discuss the bias and asymmetry controlled evolution of the gap in the Floating case, which is very much similar to the case of NISIN layout with arbitrary asymmetry of the tunnel barriers considered in Ref.~\cite{Snyman_2009}. Everywhere in the following  we neglect the inverse proximity effect, i.e. the impact of the semiconductor NW material on the superconductivity in the S shell. This approximation seems reasonable since a hard superconducting gap in the shell is very well compatible with high interface transparency in modern superconductor/semiconductor hybrids~\cite{Kjaergaard2016,Kjaergaard2017}. Also we again assume that the S-island is long enough that allows to neglect the end regions of the size on the order of the wire diameter (or the superconductor's coherence length if it's bigger) where the quasiparticle charge current in the core is converted into the supercurrent in the shell.

\subsection{Sub-gap transport: proximity effect\label{subgap_proximity}}

Sub-gap quasiparticle transport is unique to proximity NSN devices considered here and takes place for $|\varepsilon|<\Delta$ as long the superconducting gap in the S segment is finite. In this energy window all  quasiparticles reside in the proximitized normal core, where the superconducting pairing potential is absent, whereas the supercurrent flows predominantly in the superconducting shell. Unlike in all-metal devices~\cite{Keizer2006,Vercruyssen_2012}, for bias voltages $|eV|\sim\Delta$ the typical currents in NW-based devices are orders of magnitude smaller compared to the critical current of the shell. Thus a current-driven renormalization of the superconducting gap and the terms containing the gradient of the phase of the superconducting order parameter can be safely neglected and the Usadel equations greatly simplify. Furthermore, in the  the kinetic part of the Usadel equations, the energy and charge components of the non-equilibrium EED also decouple. In the absence of non-equilibrium suppression of the superconducting gap, that is in the Grounded case and in the maximally asymmetric Floating case as we explain in section~\ref{Usadel} below, the equations for the retarded Green's function remain identical to the equilibrium case and decouple from the kinetic equations. Hence, the density of states and quasiparticle diffusion coefficient in the proximitized normal core acquire the energy dependence but remain independent of the bias and coordinate along the wire (apart from the end regions of the S segment we neglect here). Exception is the general Floating case, where for high enough bias voltages $\Delta$ starts to depend on $V$ that inevitably modifies the energy dependence compared to the equilibrium case.

It is straightforward to observe the main steps of the solution of the Usadel equations for the thermal conductance. In the notations of Ref.~\cite{Keizer2006}, the energy current density in the normal core of the S segment is given by $j_\mathrm{energy} = \Pi_L\nabla f_\mathrm{long}$, where $f_\mathrm{long}$ is the longitudinal part of the EED that depends only on the coordinate along the wire and coincides with our $-F(\varepsilon,x)$. The gradients along the axes $y$ and $z$ vanish because no energy current flows transverse to the wire. $\Pi_L = \Pi_L(\varepsilon,y,z)$ takes into account the proximity effect induced energy and coordinate dependent renormalization of the thermal conductivity, which tends to zero near the superconducting interface and for $\varepsilon\rightarrow 0$ and is maximum near the center of the  core. Since $\Pi_L$ is independent of $x$ the solution~(\ref{eq:f_subgap}) for the sub-gap EED remains valid, now with the energy-dependent thermal conductance $G_\mathrm{th}(\varepsilon)\propto \int \Pi_L dydz$. This modification has an obvious consequence for the sub-gap EEDs in general case as compared to those displayed in Fig.~\ref{fig2}. While the calculated steps retain their positions $f(\varepsilon)$ between the steps acquires the energy dependence associated with that of $G_\mathrm{th}(\varepsilon)$. The actual solution of the Usadel equations which is non-universal and depends on materials, interface quality and band-structure goes beyond the scope of this work. Instead we pay attention to the main effect that the shot noise measurement allows to extract the full energy dependence of the $G_\mathrm{th}(\varepsilon)$, whatever it is, as addressed in section~\ref{Edep}.

\subsection{Impact of the non-equilibrium EED on $\Delta$}\label{Usadel}

It is well known that non-equilibrium EED generated by voltage bias results in a complex evolution of the superconducting gap.
In symmetric all-metal NSN devices~\cite{Keizer2006} the gap withstands bias voltages of about  $|eV|\approx1.4\Delta_0$, whereas in NISIN devices a finite superconducting gap is observable for $|eV|\gg\Delta_0$ given a strong asymmetry of the tunnel barriers~\cite{Snyman_2009}, where $\Delta_0$ is the $T=0$ BCS value of the gap. The bias evolution of the gap is irrelevant for our Grounded case, as well as for the maximally asymmetric Floating case, since in both these situations the S-island is ideally coupled to the reservoir and the EED remains equilibrium for arbitrary bias voltages. In the general Floating case, however, the bias and asymmetry controlled suppression of $\Delta$  accompanied by a hysteretic behaviour with bias sweeps is expected much like in the all-metal layouts~\cite{Keizer2006,Snyman_2009}.

To illustrate the main effect we numerically solved the Usadel equations together with the BCS self-consistency equation on $\Delta$~\cite{Keizer2006}. The calculations were taken in the limit of very long S-island, that corresponds to the case of bulk superconductor with a homogeneous non-equilibrium EED. This is nominally identical to a NISIN structure in the limit of vanishing Thouless energy~\cite{Snyman_2009}. In essence, we followed a slightly modified iterative approach of Ref.~\cite{Keizer2006}. First, the Green's functions were found for a certain $\Delta$ as a solution for retarded Usadel equations. For each value of the bias voltage the starting point was either $\Delta=\Delta_0$ or $\Delta<10^{-2}\Delta_0$, giving in the end two separate stable solution branches. We interpret this as a consequence of bistablility of the bias voltage characteristics observed in earlier work~\cite{Keizer2006,Snyman_2009} and relate the two solutions to the up-sweep and down-sweep branches, respectively. Second, the obtained Green's functions together with the EEDs for the same $\Delta$, Eqs.~(\ref{eq:f_subgap}-\ref{eq:f_abovegap}), were substituted in a self-consisted equation on $\Delta$. In this way the next iteration of $\Delta$ was found and the procedure repeated. We neglected the superconducting phase gradient that corresponds to the case of electric current much smaller than the critical current. We used \texttt{quad} integration from scipy (based on Fortran library QUADPACK) and typically the procedure converged after $\sim1000$~iterations.

\begin{figure}[t]
	\begin{center}
		\includegraphics[width=1\columnwidth]{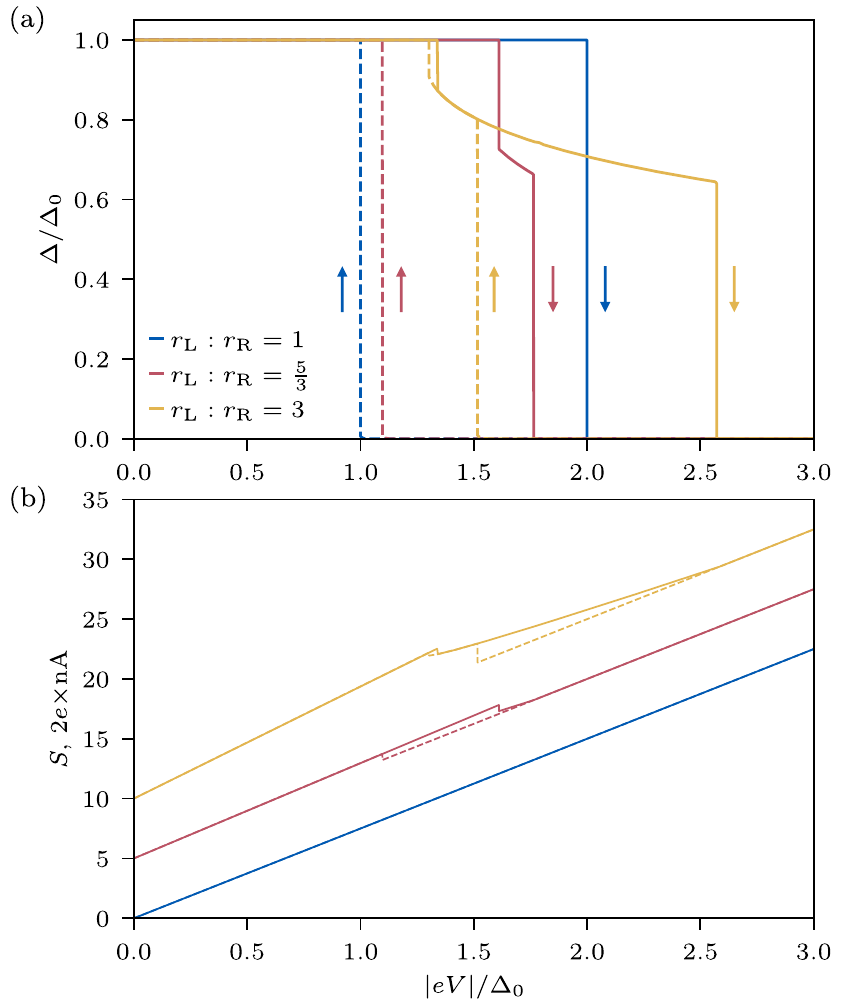}
	\end{center}
	\caption{Non-equilibrium suppression of the superconducting gap and its manifestation in shot noise in the Floating case. (a) Bias evolution of the superconducting gap for several ratios  $r_\mathrm{L}:r_\mathrm{R}$ given in the legend. Solid and dashed lines correspond, respectively, to the sweep-up and sweep-down solution branches. (b) Shot noise for the same resistance ratios and the total device resistance of $8\,\mathrm{k\Omega}$. Line styles are the same as in panel (b), the data is vertically offset for clarity with the asymmetry increasing from the lower to the upper curve. For this data the sub-gap thermal conductance is assumed negligible $G_\mathrm{th}(|\varepsilon|<\Delta)=0$.}
	\label{fig_Usadel}
\end{figure}

The obtained bias dependencies of $\Delta$ are shown in Fig.~\ref{fig_Usadel}a for several device asymmetries determined by the resistance ratio $r_\mathrm{L}:r_\mathrm{R}$ and fixed total resistance of $r_\mathrm{L}+r_\mathrm{R}=8\,\mathrm{k\Omega}$. In a symmetric device, the gap value is found to switch between $\Delta=\Delta_0$ and $\Delta=0$ and the switch position depends on the sweep direction indicated by arrows. This behaviour is qualitatively consistent with the numerical data of Ref.~\cite{Keizer2006} and coincides with the analytical result of Ref.~\cite{Snyman_2009} in the limit of vanishing Thouless energy. The key impact of the device asymmetry is the possibility of the intermediate stable gap values  $\Delta_0>\Delta>0$, as demonstrated for two other values of the asymmetry in Fig.~\ref{fig_Usadel}a. For $r_\mathrm{L}\neq r_\mathrm{R}$ we always find a region of bias voltages where the intermediate gap value is stabilized for the sweep-up direction, and sometimes for the sweep-down direction (e.g, for the resistance ratio of 3). Similarly to the symmetric case, the bias dependencies exhibit bistability, however the finite gap values can exist up to arbitrarily large bias voltages, given the arbitrary large asymmetry. All these findings are consistent with the detailed studies of the all-metal NISIN structures in Ref.~\cite{Snyman_2009}.

Comparatively new results correspond to the shot noise behaviour in our NSN devices. While the details of the noise calculations are presented later in section~\ref{noise_section} in the context of sub-gap thermal conductance, it is convenient to show the shot noise data corresponding to the non-equilibrium evolution of the superconducting gap here. Fig.~\ref{fig_Usadel}b demonstrates the calculated current noise spectral density in the Floating case for the same three values of the asymmetry. In order to demonstrate solely the gap related behaviour we have taken $G_\mathrm{th} =0$ in this panel, so that the sub-gap quasiparticle transport is forbidden. The bath temperature is zero and the data are vertically offset for clarity. In the symmetric case (lower curve), the shot noise is insensitive to $\Delta$ and always coincides with the universal value in normal diffusive conductors. This is related to the fact that the EED given by Eqs.~(\ref{eq:f_subgap}-\ref{eq:f_abovegap}) has the same double-step shape as in the middle of a normal diffusive conductor~\cite{Nagaev1992}  regardless of both the gap value and the thermal conductance. By contrast, for $r_\mathrm{L}\neq r_\mathrm{R}$ the shot noise exhibits a jump at a lower bias and a kink at a higher bias for the sweep-up and two jumps for sweep-down bias dependencies, as evident from the figure. The positions of these features coincide with the steps on the corresponding dependencies $\Delta(V)$ in Fig.~\ref{fig_Usadel}a.

The results of this section can be summarized in the following two conclusions. First, the effect of the non-equilibrium gap suppression in the Floating case can complicate the analysis of the sub-gap thermal conductance via shot noise. The reason is that the value of the gap is directly related to the energy dependence of $G_\mathrm{th}(\varepsilon)$ via the proximity effect. The stronger the asymmetry the less pronounced is the non-equilibrium, completely absent only in the marginal case of vanishing resistance in one of the N segments. Second, an imperfect contact to the superconducting shell in NW-based devices via a semiconducting core or a tunnel junction has an inevitable consequence for the superconducting gap due to non-equilibrium effects at high enough bias voltages~\cite{Keizer2006,Snyman_2009}, that may be a considerable effect in modern NW-based devices.

\section{Shot noise\label{noise_section}}

Knowing the EEDs on the boundaries of the S segment, Eqs.~(\ref{eq:f_subgap}-\ref{eq:f_abovegap}), one finds the $f(\varepsilon, x)$ and the local noise temperature $T_\mathrm{N}(x)$. The spectral density of the spontaneous current fluctuations in the normal segments of the NW is then calculated using the semiclassical solution for diffusive conductors~\cite{Nagaev1992}:
\begin{align}
	T_\mathrm{N}(x) &= \frac{1}{k_\mathrm{B}} \int_{-\infty}^{+\infty} f(\varepsilon,x)(1-f(\varepsilon,x)) d \varepsilon \nonumber\\
	S_\mathrm{L} &= \frac{4 k_\mathrm{B}}{r_\mathrm{L}} \int_{x_\mathrm{L}}^{x_1} T_\mathrm{N}(x) d x \label{nagaev_noise}\\
	S_\mathrm{R} &= \frac{4 k_\mathrm{B}}{r_\mathrm{R}} \int_{x_2}^{x_\mathrm{R}} T_\mathrm{N}(x) d x.\nonumber
	\end{align}

A separate measurement of  the fluctuations $S_\mathrm{L}$ and $S_\mathrm{R}$ is possible only in the Grounded case~\cite{Denisov2020}. In the Floating case the normal segments are connected in series and the resulting current fluctuation is~\cite{Beenakker1992}:
\begin{align*}
	S = \frac{S_\mathrm{L}r_\mathrm{L}^2 + S_\mathrm{R}r_\mathrm{R}^2}{(r_\mathrm{L} + r_\mathrm{R})^2}.
\end{align*}

Note that in the last equation the contribution of the S segment to the measured shot noise is absent thanks to its negligible resistance. Nevertheless, the role of the thermal conductance in the S segment is decisive, since it is $G_\mathrm{th}$ that determines the non-equilibrium EEDs on the boundaries of S segment. In the following we first discuss the results for the case of $G_\mathrm{th}(\varepsilon)=const$. Although this overly simplified limit of negligible proximity effect is hard to realize in contemporary NW-based devices, we find it useful for the demonstration of the main features in the shot noise behaviour. Later on we discuss how the results change in the general case of the arbitrary energy dependent thermal conductance and take a closer look at a few interesting cases of the dependence $G_\mathrm{th}(\varepsilon)$.

\subsection{Results neglecting the energy dependence}

\begin{figure}[tb]
	\begin{center}
		\includegraphics[width=0.9\columnwidth]{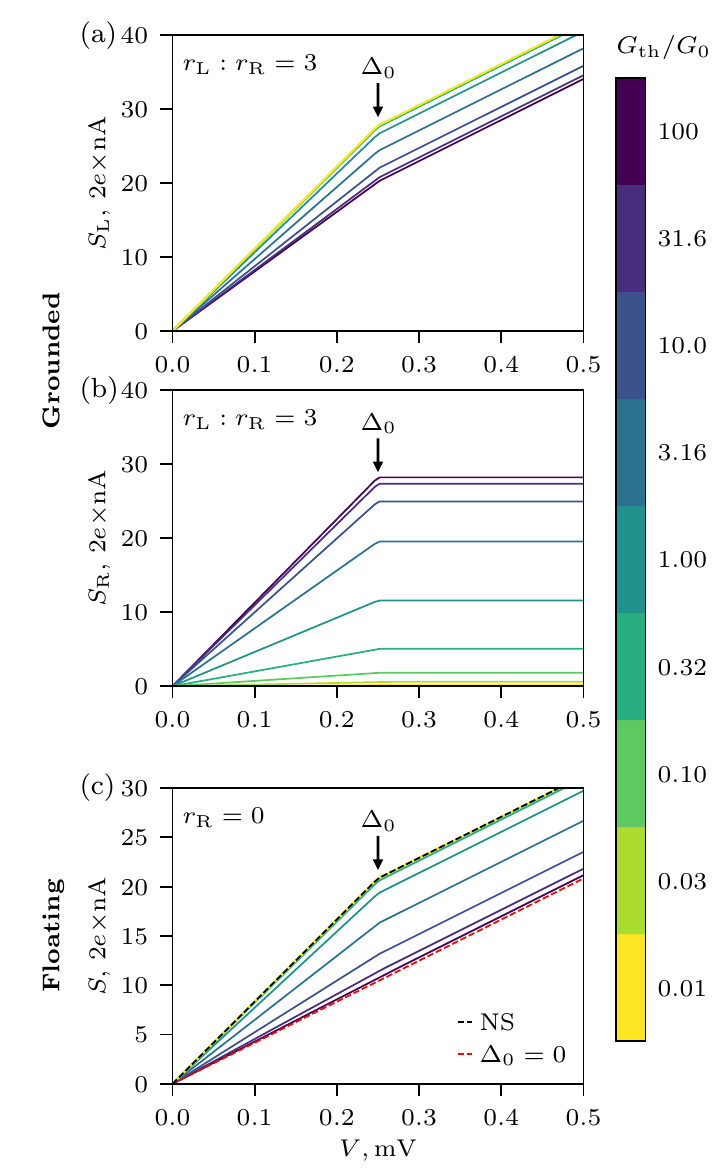}
	\end{center}
	\caption{Zero temperature shot noise evaluated at various $G_\mathrm{th}$ expressed in units of $G_\mathrm{0}=2e^2/h$. (a-b)~Current fluctuations of the biased ($S_\mathrm{L}$) and unbiased ($S_\mathrm{R}$) normal segments in the Grounded geometry, $r_\mathrm{L}:r_\mathrm{R} = 4$. (c)~Current fluctuations in the maximally asymmetric Floating case. Red dashed is the guide line for ordinary metallic diffusive wire, $F=1/3$. Black dashed line represents the  dirty NS junction without the sub-gap states \cite{Nagaev2001}. In (a-b) the resistances of the normal segments are $r_\mathrm{L}=6\,\mathrm{k}\Omega$ and $r_\mathrm{R}=2\,\mathrm{k}\Omega$. In (c) $r_\mathrm{L}=8\,\mathrm{k}\Omega$ and $r_\mathrm{R}=0$.}
	\label{fig3}
\end{figure}

\begin{center}
	\textit{Grounded}
\end{center}

Using the notations ${\theta = [1+G_\mathrm{th}(r_\mathrm{L}+r_\mathrm{R})]^{-1}}$ and ${\alpha=r_\mathrm{L}/(r_\mathrm{L}+r_\mathrm{R})}$ we express the general solutions for the $S_\mathrm{L}$ and $S_\mathrm{R}$ in the Grounded case  in the zero temperature limit as follows:
\begin{align}
	&S_\mathrm{L} = \frac{2eV}{3r_\mathrm{L}} + \frac{2e}{3r_\mathrm{L}}\Xi_\mathrm{L}
	\begin{cases}
		V, & |eV| < \Delta_0 \\
		\Delta_0, & |eV| \ge \Delta_0 \label{eq:SL}
	\end{cases}\\	
	&S_\mathrm{R} = \frac{2e}{3r_\mathrm{L}}\Xi_\mathrm{R}
	\begin{cases}
		V, &|eV| < \Delta_0 \\
		\Delta_0, & |eV| \ge \Delta_0 \label{eq:SR}
	\end{cases} \\
	&\Xi_\mathrm{L}\left(\alpha, \theta\right) = 1-(\alpha-\alpha\theta)^2, \nonumber \\
	&\Xi_\mathrm{R}\left(\alpha, \theta\right) = \alpha (1 - \theta) (2 + \theta + \alpha (1 - \theta)). \nonumber
\end{align}

Representative results for the shot noise spectral density are plotted in \reffig{fig3}(a) and \reffig{fig3}(b). Here, the $G_\mathrm{th}$ is energy independent and the correspondence between the line colours and the $G_\mathrm{th}$ values is shown on the nearby colour bar. Both current fluctuations $S_\mathrm{L}$ and $S_\mathrm{R}$ demonstrate a kink at $|eV|=\Delta_0$, when the voltage applied on the left N terminal meet the superconducting gap edge. The effect of finite sub-gap thermal conductance of the S segment on the $S_\mathrm{L}$ and $S_\mathrm{R}$ is different. The noise of the biased normal segment $S_\mathrm{L}$ is maximum if the thermal transport is suppressed and diminishes at increasing $G_\mathrm{th}$. By contrast, the noise of the unbiased normal segment $S_\mathrm{R}$, that originates from the quasiparticles transmitted via the proximity region, is only observable at $G_\mathrm{th}>0$. Such an impact of the thermal conductance on the shot noise was observed in a recent experiment~\cite{Denisov2020} and is easy to understand qualitatively. Increasing the thermal conductance allows more sub-gap quasiparticles to relax in the right N terminal. This is analogous to a cooling down of the biased left N segment and a warming up of the unbiased right N segment, with obvious consequences for the noise. Note that $S_\mathrm{R}$ saturates at  $|eV|>\Delta_0$, since above the gap the  quasiparticles sink in the grounded S terminal.

\begin{center}
	\textit{Floating}
\end{center}

Using the same notations, and assuming  without the loss of generality that $\alpha \ge \frac{1}{2}$, we express the results for the shot noise in the Floating case as follows:
\begin{align}
	&\begin{aligned}
	S &= \frac{2eV}{3R} \\
	  &+ \frac{2e}{3R}\Gamma
	\begin{cases}
		(2\alpha-1)V, & \alpha |eV| < \Delta\\
		-(1-\alpha)V + \Delta, & else\\
		0, & (1-\alpha)|eV| \ge \Delta \label{eq:S}
	\end{cases}
	\end{aligned} \\
	&\Gamma\left(\alpha, \theta\right) = 1 - 3\alpha^2(\theta -1)\theta + \alpha\left(3\theta^2-1\right)-\theta (\theta +1), \nonumber
\end{align}
where $R=r_\mathrm{L}+r_\mathrm{R}$ is end-to-end resistance of the device.

These equations are applicable to the Floating case of arbitrary asymmetry. As discussed in section~\ref{Usadel}, the value of $\Delta$ generally exhibits a bistable evolution with the bias voltage, which manifests itself in shot noise, see the data of Fig.~\ref{fig_Usadel}b. Those data were calculated with the Eqs.~(\ref{eq:S}) for $\theta=1$ (i.e. $G_\mathrm{th}=0$). In order to disentangle the effects of the thermal conductance and gap evolution we address in \reffig{fig3}(c) the limiting case of the maximally asymmetric Floating device $r_\mathrm{R}= 0$, for which  $\Delta=\Delta_0$ regardless of the bias voltage. In this case we observe a single kink at the usual position $|eV|=\Delta_0$ and the $S(V)$ dependence interpolates between the well known limits of diffusive normal and NS cases depending on $G_\mathrm{th}$.

\subsection{Energy dependent thermal conductance \label{Edep}}

\begin{figure}[tb]
	\begin{center}
		\includegraphics[width=1\columnwidth]{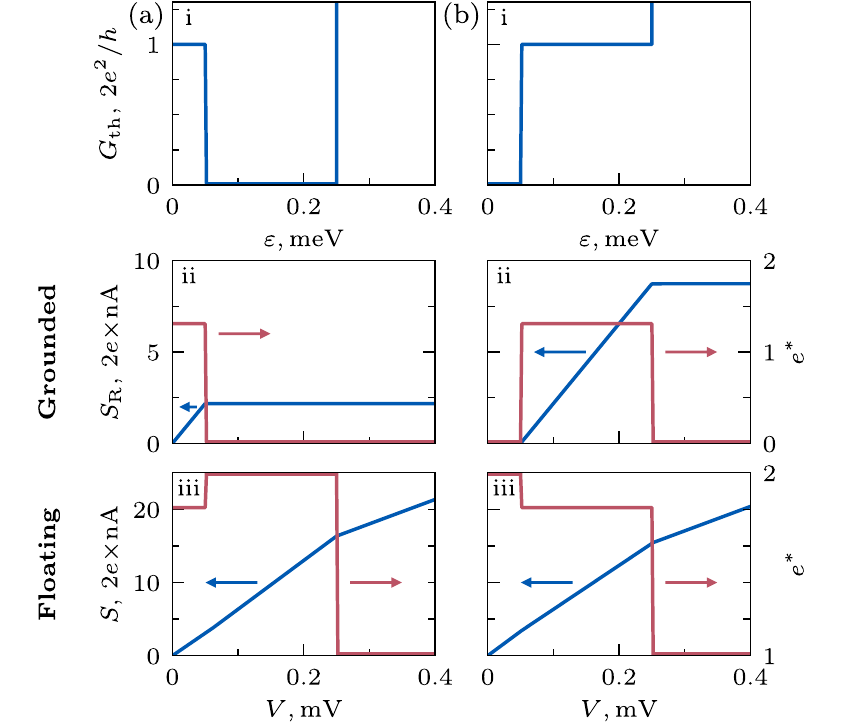}
	\end{center}
	\caption{Measurement of energy dependent $G_\mathrm{th}(\varepsilon)$ via shot noise. (i)~Stepwise $G_\mathrm{th}(\varepsilon)$ functions imitating the (a)~topological phase transition and (b)~hard gap $\Delta_1 = 50\,\mathrm{\upmu eV}$. (ii)~Current fluctuations of the unbiased segment and $e^*$ in the Grounded setup. (iii)~Current fluctuations and $e^*$ in the Floating setup.}
	\label{fig4}
\end{figure}

In previous sections we treated the problem for the energy independent thermal conductance. Thanks to the absence of the energy relaxation, however, the  shot noise measurement provides access to a full energy dependence of $G_\mathrm{th}(\varepsilon)$, which may result from the energy dependence of the quasiparticle density of states as a consequence of the superconducting proximity effect. This information is contained in a slope $\partial S/\partial I$ and is conveniently expressed  in terms of a renormalized effective charge $e^*$ as follows:
\begin{align}
	&\text{Grounded:} \nonumber\\
 	\begin{split}
		|eV|<\Delta_0\!:\, e^*& \equiv \frac{1}{2eF} \frac{\partial S_\mathrm{R}}{\partial I_\mathrm{L}} =\Xi_\mathrm{L}\left(\alpha, \theta\right) \\
		|eV|>\Delta_0\!:\, e^* &=0
		\label{eq:estar_gnd}
	\end{split}\\
	&\text{Floating},\,r_\mathrm{R}\rightarrow0\!: \nonumber\\
	\begin{split}
		|eV|<\Delta_0\!:\, e^* & \equiv \frac{1}{2eF} \frac{\partial S}{\partial I} = 1+\Gamma\left(1, \theta\right) \\
		|eV|>\Delta_0\!:\, e^* & = 1,
		\label{eq:estar_floating}
	\end{split}
\end{align}
where $F = 1/3$ is the universal value of the Fano factor in  metallic diffusive conductors. The energy dependence of the sub-gap thermal conductance enters the expression for the effective charge via ${\theta(\varepsilon) = [1+G_\mathrm{th}(\varepsilon)(r_\mathrm{L}+r_\mathrm{R})]^{-1}}$. Similarly to the Grounded case~(\ref{eq:estar_gnd}), the limit $r_\mathrm{R} \rightarrow 0\,(\alpha=1)$ of the Floating case provides the direct relation between $e^*$ and $G_{\text{th}}(eV)$ via expression~(\ref{eq:estar_floating}). In contrast, a generic Floating case with $\alpha<1$ is less convenient since the effective charge is dependent on both $G_\mathrm{th}(\alpha eV)$ and $G_\mathrm{th}(eV - \alpha eV)$, not to mention the bias dependent renormalization of $\Delta$ and bistability discussed in section~\ref{Usadel}. The corresponding bulky expressions for the $e^*$ are offloaded to the Appendix~\ref{appendix_2}.

In the following we illustrate these results for the simplest realizations of $G_\mathrm{th}(\varepsilon)$, see \reffig{fig4} (row i). $G_\mathrm{th}(\varepsilon)$ is assumed to be a stepwise function:
\begin{align*}
	G_\mathrm{th}(\varepsilon) = 
	\begin{cases}
		G_1, &  |\varepsilon| < \Delta_1 \\
		G_2, &  \Delta_1 \le |\varepsilon| < \Delta_0, \\
			\end{cases}
\end{align*}
which imitates a one dimensional wire for the case of a topological phase transition $(G_1, G_2) = (2e^2/h, 0)$~\cite{Akhmerov2011} and for the case of a hard superconducting gap $(G_1, G_2) = (0, 2e^2/h)$ of width $\Delta_1 < \Delta_0$~\cite{Chang2015, Kjaergaard2016}.

In \reffig{fig4} we plot the bias dependencies of the shot noise and $e^*$ for these two situations, respectively, in column~(a) and column~(b). Here, $r_\mathrm{L}=10\,\mathrm{k}\Omega$ and $r_\mathrm{R}\rightarrow0$ so that $\Delta=\Delta_0$ is independent of the bias. For both device layouts the shot noise spectral density shows a kink each time the $G_\mathrm{th}$ changes abruptly. Consequently, the shape of the bias dependence of the $e^*$ mimics the energy dependence of the thermal conductance, cf. the panels~(ii) and~(iii) with the corresponding panels~(i) in \reffig{fig4}. Two key differences between the Grounded and Floating cases are evident. First, in the Grounded case the $e^*$ increases as function of $G_\mathrm{th}$, whereas in the Floating case the dependence is opposite. Second, in the  Grounded case the effective charge varies between~0 and~3, $e^*=0$ corresponding to $G_\mathrm{th}=0$, whereas in the Floating case  $1\leq e^*\leq2$, $e^*=2$ corresponding to $G_\mathrm{th}=0$.

\section{Discussion}

In this section we briefly summarize our results and their implications for the ongoing research in hybrid devices. 

One important message is a possibility to probe the sub-gap quasiparticle transport in NSN devices by means of the shot noise response, which was initially proposed for Majorana devices in Ref.~\cite{Akhmerov2011} and generalized here for the diffusive case with both Grounded and Floating S-island. The underlying physics is straightforward --- electric current at least in one arm of the NSN device gives rise to the non-equilibrium EED in both arms and results in spontaneous fluctuations of the electric current owing to a stochastic nature of the quasiparticle transport. In the semiclassical framework~\cite{Nagaev1992}, according to the Eqs.~(\ref{eq:f_subgap},\ref{eq:f_abovegap}), the non-equilibrium EED is determined by just two quantities --- the superconducting gap and the energy dependent thermal conductance. As discussed in section~\ref{subgap_proximity}, in the Grounded and maximally asymmetric Floating device layouts the S-island is effectively coupled to a quasiparticle reservoir, so that both $\Delta$ and $G_\mathrm{th}(\varepsilon)$ are independent of the applied bias voltage/current. This explains the remarkable simplicity of our final shot noise expressions  in Eqs.~(\ref{eq:estar_gnd}) and (\ref{eq:estar_floating}). 

By contrast, the final results in the general Floating case are not just bulky, see the Appendix~\ref{appendix_2}, but also require a self-consistent solution of the Usadel equations to find $\Delta$ and $G_\mathrm{th}(\varepsilon)$ at each given value of the bias voltage (and sweep direction), as we discussed and exemplified  in section~\ref{Usadel}. Obviously, such a non-universality complicates the analysis of the shot noise experiments in terms of the sub-gap thermal conductance. This is the second important message of our work --- being decoupled from a quasiparticle reservoir the superconductor takes over the key role in the non-equilibrium problem, and can no longer be treated as a fixed boundary condition for the neighbouring semiconductor. Note that such a decoupling may occur not only by purpose~\cite{Vaitiekenas2018}, but also unintentionally in the form of a finite contact resistance to the superconducting shell~\cite{stampfer2021}. The connection between the non-equilibrium superconductivity~\cite{Keizer2006,Snyman_2009,Catelani_2010,Vercruyssen_2012,Serbyn_2013} and the physics of hybrid semiconductor-superconductor devices remains, to our best knowledge, unexplored. We hope that our work will motivate further interest in this direction within the community. 

Finally, we intended to popularize a more general message that the shot noise is a valuable research tool in superconducting proximity devices. Recent experiments~\cite{Denisov2020,Denisov2021} demonstrate that a full understanding of the non-local quasiparticle signals in NSN devices of Grounded type requires a combination of charge transport measurement and shot noise measurement, the latter substituting the thermal measurement. Already in a few hundred nanometer long devices, the transmitted quasiparticle flux is charge-neutral that gives noise measurement a primary role~\cite{Denisov2021}. By contrast, the non-local charge response is a secondary thermoelectric-like effect~\cite{Denisov2020} caused by the asymmetry of the energy dependence of the spectral conductance with respect to $\varepsilon=0$. In our semiclassical treatment we have neglected any such asymmetry thus focusing on the main effect of the heat transport in the proximitized region.

In summary, we analyzed how a finite sub-gap thermal conductance of a superconducting proximity region and non-equilibrium suppression of superconductivity are manifested in the shot noise of diffusive NSN structures. Two possible device layouts --- three-terminal with a grounded superconducting island and two-terminal with a floating superconducting island --- permit a direct measurement of the energy dependence $G_\mathrm{th}(\varepsilon)$ from the bias dependence of the shot noise. The Floating case may be attractive for the experimental realization in semiconducting nanowires with a thin superconducting shell that maybe technically challenging to contact. In this case the device asymmetry is crucial in the shot noise experiment, that can be engineered by means of the structure design and/or local gating. However, an imperfect grounding of the superconducting island makes the Floating case susceptible to the well-known effects of non-equilibrium suppression of the superconducting gap and its bistable behaviour at high enough bias voltages. These effects are also observable in shot noise the only exception being the limiting case of maximal asymmetry. Applicable to diffusive multi-mode wire structures our results may also serve as a qualitative starting point to understand the shot noise response in open one dimensional Majorana devices.

\section{Acknowledgments}

We thank A.O.~Denisov, K.E.~Nagaev, T.M.~Klapwijk and, particularly, E.S.~Tikhonov for valuable discussions. We are grateful to Y.V.~Fominov for the insights about the Usadel theory. This work was mainly supported by the RSF project 19-12-00326. The calculation of the finite temperature shot noise expression in Appendix~\ref{appendix_1} was performed under the state task of ISSP RAS.

\appendix

\section{\label{appendix_1}\uppercase{The Grounded geometry. Finite temperature}}

Here we provide the finite temperature expression for shot noise of the unbiased normal segment in the Grounded geometry. We use $e, k_B=1, \gamma=r_\mathrm{R} /\left(r_\mathrm{L}+r_\mathrm{R} + 1/G_\mathrm{th}\right)$.
\begin{widetext}
\begin{align*}
S_\mathrm{R} = &\frac{4 T}{ r_\mathrm{R}} + \frac{2 T}{3 r_\mathrm{R}} \gamma \left\{ \left[\gamma + 3 (2 - \gamma) \coth ^2\left(\frac{V}{2 T}\right)\right]\tanh \left(\frac{V}{2 T}\right) \artanh \left[\tanh \left(\frac{\Delta }{2 T}\right) \tanh \left(\frac{V}{2 T}\right)\right] + \right.\\
&\left.\frac{1}{2} \left[\gamma - 3 (2 - \gamma) \cosh \left(\frac{\Delta}{T}\right) - 2 (3 - \gamma) \cosh \left(\frac{V}{T}\right)\right] \tanh \left(\frac{\Delta }{2 T}\right) \sech\left(\frac{V-\Delta }{2 T}\right) \sech\left(\frac{\Delta +V}{2 T}\right) \right\}
\end{align*}
\end{widetext}

\section{\label{appendix_2}\uppercase{General result in the Floating geometry}}

In the general case (arbitrary $r_\mathrm{L}, r_\mathrm{R}$ and energy dependent $G_\mathrm{th}(\varepsilon)$) shot noise in the Floating layout exhibits 2 kinks (see \reffig{fig3}(c)). Here we write the expression for $e^*$ defined in \refeq{eq:estar_floating}, where $\theta$ contains $G_\mathrm{th}(\varepsilon)$. We remind the reader, that the following expressions should be supplemented with the calculation of the bias-dependent suppression of the superconducting gap due to non-equilibrium effects.
\begin{widetext}
\begin{align*}
  e^* = 
    \begin{cases}
      \alpha (1 - \alpha) + \alpha \left(3 \alpha ^2 - 1\right) \theta (\alpha V) - \alpha \left(3 \alpha ^2 - 3 \alpha + 1\right) \theta ^2 (\alpha V) +\\
      \alpha (2 - \alpha) + \left(3 \alpha ^3 - 3 \alpha ^2 - \alpha + 1\right) \theta (V (1 - \alpha)) + \left(-3 \alpha ^3 + 6 \alpha ^2 - 4 \alpha + 1\right) \theta ^2 (V (1 - \alpha)), & \alpha V < \Delta\\
      \alpha (2 - \alpha) + \left(3 \alpha ^3 - 3 \alpha ^2 - \alpha + 1\right) \theta (V (1-\alpha)) + \left(-3 \alpha ^3 + 6 \alpha ^2 - 4 \alpha + 1\right) \theta ^2 (V (1 - \alpha)), & else\\
      1, & (1 - \alpha) V  \ge \Delta\\
    \end{cases}
\end{align*}
\end{widetext}


%

\end{document}